# Why Did Life Emerge?

Arto Annila* and Erkki Annila[#]

*Department of Physics, Institute of Biotechnology and Department of Biosciences, POB 64, University of Helsinki, Finland, [#]Finnish Forest Research Institute, Finland

**Abstract**

Many mechanisms, functions and structures of life have been unraveled. However, the fundamental driving force that propelled chemical evolution and led to life has remained obscure. The $2^{nd}$ law of thermodynamics, written as an equation of motion, reveals that elemental abiotic matter evolves from the equilibrium via chemical reactions that couple to external energy toward complex biotic non-equilibrium systems. Each time a new mechanism of energy transduction emerges, *e.g.*, by random variation in syntheses, evolution prompts by punctuation and settles to a stasis when the accessed free energy has been consumed. The evolutionary course toward an increasingly larger energy transduction system accumulates a diversity of energy transduction mechanisms, *i.e.*, species. The rate of entropy increase is identified as the fitness criterion among the diverse mechanisms which places the theory of evolution by natural selection on the fundamental thermodynamic principle with no demarcation line between inanimate and animate.

## 1. Introduction

The theory of evolution by natural selection[1] pictures how biodiversity[2] has cumulated. Fossil records and similarity among biological macromolecules are rationalized by projecting back in time from the contemporary branches of life along paths that merge over and over again into common ancestors[3]. When descending down to the epoch of chemical evolution[4,5], devoid of genetic material and apparent mechanisms of replication, it is unclear how natural selection operates[6,7,8] on matter and yields functional structures and hierarchical organizations that are characteristics of life.

The basic question, *why* matter evolved from inanimate to animate, is addressed in this study using the theory of evolution by natural selection that was recently formulated in thermodynamic terms[9]. In nature many phenomena follow the $2^{nd}$ law of thermodynamics known also as the principle of increasing entropy[10]. The law, as it was given by Carnot, is simple: an energy difference is a motive force[11]. For example, heat flows from hot to cold and molecules diffuse from high to low concentration. Energy flows also in chemical reactions that transform compounds to other compounds to diminish chemical potential energy differences. Eventually a stationary state without energy gradients is reached. For example, the chemical equilibrium[12,13] corresponds to the most probable distribution of reactants and products. In general, all processes that level potential energy gradients are referred to as *natural processes*[14].

According to thermodynamics evolution in its entirety is also a natural process driven by the universal tendency to diminish differences among energy densities. Although the quest for higher entropy has for long been understood as the primus motor of evolution and as the emergent motive for orderly mechanisms and hierarchical organizations[15,16,17,18,19,20,21,22,23,24,25,26], it nevertheless seems that the 2$^{nd}$ law has not acquired unanimous recognition as the profound principle that governs also processes that we refer to as living. The physical basis of the entropy law was recently strengthened when it was derived from probability considerations and formulated as an equation of motion[9]. Now it is possible to deduce unmistakably where a system under an influx of external energy is on its way. In particular it can be understood, what is happening when external energy from Sun couples to numerous chemical reactions that distribute matter on Earth.

The recently derived equation of evolution[9] has already been used to account for the emergence of chirality consensus and other standards of life[27] as well as to tackle the puzzle of large amounts of non-expressed DNA in eukaryotes[28]. Furthermore, skewed population distributions that are ubiquitous characteristics of plant and animal populations just as gene lengths and their cumulative curves, *e.g.*, species-area relationships have been shown to be consequences of the 2$^{nd}$ law[29,30]. Also the global homeostatic characteristics that were articulated by Gaia theory[31], have been placed on the same thermodynamic foundation[32]. Moreover, the ubiquitous imperative to disperse energy has been associated with the principle of least action to describe flows of energy. The flows direct down along the steepest gradients, equivalent to the shortest paths, and flatten the manifold of energy densities[33].

In this study, evolution, on all length scales and at all times, is considered to display the ubiquitous principle of energy dispersal. The subsequent thermodynamic analysis does not bring forward essentially novel thoughts but communicates the simple physical basis that underlies the earlier reasoning about emergence of life, rise of complexity and courses to hierarchical organizations. It is emphasized the study does not aim to expose any particular locus or moment in time or precise primordial conditions from which life sprang up. In fact, thermodynamics gives no special attributes to living systems but describes all matter as compounds, *i.e.* heterogeneous substances[12], at large entities. To recognize energy gradients as evolutionary forces paves the way for understanding why life emerged.

## 2. On the entropy concept

The adopted view of entropy, *i.e.*, entropy increases when energy gradients diminish, is briefly contrasted with other notions associated with the entropy concept. The standpoint is traditional thermodynamic because an energy gradient is understood as a motive force but the equation of motion has been obtained from the statistical probability calculation. In contrast the informational entropy defined mathematically by Shannon[34] does not explicitly recognize probability as a physical motive[35]. Even without explicit energetic terms it is possible to deduce mathematically, *e.g.*, using Lagrange multipliers, the maximum entropy state because per definition at the stationary state the energy differences, *i.e.*, the driving forces have vanished. However, when using informational entropy as such the evolutionary course itself that arrives at the stationary state remains unclear. The maximum entropy principle formulated by Jaynes[36] builds on the abstract informational entropy but aims at finding the paths that lead to increasingly more probable states. These optimal paths are associated with the steepest ascents and found by imposing constraints. The resulting principle of maximum entropy production for non-equilibrium stationary states[24] parallels the thinking in this study. However, the imposed constraints do not substitute for the adopted formalism that describes mutually interdependent entities in energetic terms. The diminishing energy density differences will without further guidance direct the course along the shortest paths that are equivalent to the steepest descents in the energy landscape[33]. Furthermore, it is important to realize that the driving forces keep changing due to the motion that, in turn, affects the forces. In other words, the trajectory of evolution is non-deterministic. The course of a system is not predetermined by the initial conditions or constraints because the system is changing irreversibly either by acquiring or loosing energy.

The adopted standpoint makes no principal distinction between the concepts of non-equilibrium and equilibrium. Typically systems that grow in their energy density are referred to as animate whereas those that shrink are regarded mostly as inanimate. However, in both cases the principle to diminish gradients is the same. Both animate and inanimate systems aim at stationary states governed by the high-energy and low-energy surroundings, respectively. Customarily the resulting high-energy animate state is referred to as the non-equilibrium whereas the low-energy inanimate state is referred to as the equilibrium state. Here the stationary state concept is preferred for both systems to denote the state when there is an energy balance between the system and its surroundings irrespective whether the surroundings is high or low in energy density. It is, of course, somewhat of a subjective decision how one wishes to label some entities as being parts of the system and others as being parts of the surroundings. However the choice is of no consequence when using the adopted formalism. Entropy of the system just as entropy of its surroundings will increase as mutual differences in energy are leveling off.

Finally it is emphasized that the adopted standpoint does not associate high entropy with high disorder[37]. Certainly many animate processes are driven to orderly functional structures to attain stationary states in their high-energy surroundings just as many inanimate processes are driven to disintegrate to disordered aggregates to attain stationary states in their low-energy surroundings. However, order or disorder is a consequence of energy dispersal, not an end to itself or a motive force.

## 3. Evolution as a probable process

Consequences of thermodynamics to the emergence of life are perhaps best exemplified by considering a primordial pool[1,5] that contains some basic compounds. The compounds make a chemical system by reacting with each other and coupling to an external source of energy, *e.g.*, to high-energy radiation from Sun. The system is an energy transduction network that disperses energy influx via chemical reactions among all compounds. Obviously the particular compounds that happen to be in the pool are very important for conceivable chemistry but to elucidate the general driving force that propels evolution no presumptions are made about the ingredients. In other words, the important mechanistic questions of *how* life came about are not addressed in this study but the driving force, *i.e.*, the cause *why* life emerged is clarified.

It is perhaps a common thought but a misconception that chemical reactions would be random without any preferred direction. Reactions do take the direction of decreasing free energy which is equivalent to increasing entropy, *i.e.*, the basic maxim of chemical thermodynamics. This is also the natural direction taken during chemical evolution. The motion down along energy gradients can be pictured as a sequence of steps where the system moves via chemical reactions from one distribution of primordial compounds to another in the quest for attaining a stationary state in the high-energy influx. To learn about the probable direction of motion, the plausible states, *i.e.*, distributions of compounds (entities) in numbers $N_j$ are compared by entropy[9]

$$S = R\ln P = \frac{1}{T}\sum_{j=1} N_j \left( \sum_k \mu_k + \Delta Q_{jk} - \mu_j + RT \right) \qquad (1)$$

where $\mu_k/RT = \ln[N_k\exp(G_k/RT)]$ denotes chemical potential of substrates and $\mu_j$ of products. The average energy $RT$ concept is meaningful when the system is sufficiently statistic[38]. According to Eq. 1 entropy $S$ is a logarithmic probability measure of the energy dispersal. When energy $\Delta Q_{jk}$ from the surroundings couples to a reaction, it will add to the substrate chemical potent $\mu_k$ and raise it by $\Delta Q_{jk}$ to turn the energy flow from the excited substrate potential $\mu_k+\Delta Q_{jk}$ *downhill* toward the product potential $\mu_j$ and power the endoergic reaction ($\mu_k + \Delta Q_{jk} > \mu_j$). Without the external energy the flow would be from $\mu_j$ to $\mu_k$, thus in the opposite exergic direction but also then downhill. Thermal excess of energy produced by the reaction is dissipated from the system ultimately to the cold space. Alternatively reactions may be powered by an influx of high-$\mu$ matter (*e.g.* food) that is consumed in coupled exoergic reactions to drive endoergic reactions. The resulting low-$\mu$ matter excess (*e.g.* excrement) is discarded from the system. Thus the thermodynamic formula (Eq. 1) speaks about mundane matters in the terms of physical chemistry. The value of the general expression of entropy is that it serves to describe concisely diverse energy transduction systems at various levels of hierarchy. For a particular system detailed knowledge of constituents, *e.g.*, concentrations $N_j$, Gibbs free energy $G_j$, influx $\Delta Q_{jk}$ and possible *jk*-reactions, can be given to calculate entropy using Eq. 1.

**4. The fitness criterion**

The primordial pool contains at any given moment a distribution of compounds. A reaction that turns $N_k$ to $N_j$ (or *vice versa*) will alter the distribution. The resulting distribution can be compared with the initial one by Eq. 1 to deduce if the particular reaction changed the distribution to a more probable one. Thus for any given initial state it can deduced where the chemical system is most likely to be on its way via chemical reactions. To infer the probable course of evolution the time derivative of Eq. 1 gives the 2$^{nd}$ law of thermodynamics as an equation of motion[9]

$$\frac{dS}{dt} = \sum_{j=1} \frac{dS}{dN_j}\frac{dN_j}{dt} = \frac{1}{T}\sum_{j=1}\frac{dN_j}{dt}\left( \sum_k \mu_k + \Delta Q_{jk} - \mu_j \right) = \frac{1}{T}\sum_{j=1} v_j A_j \geq 0 \qquad (2)$$

where the velocity of a reaction $v_j = dN_j/dt$. The notation is concise but it includes numerous chemical reactions that eventually result in biological functions. The potential energy difference that drives the reaction is known also as free energy or exergy or affinity[14] $A_j = \Sigma\mu_k+\Delta Q_{jk}-\mu_j$. Importantly $A_j$ includes also the energy influx. When $A_j > 0$, there is free energy to increase the concentration (or population) $N_j$ of molecular (or plant and animal) species $j$. When $A_j < 0$, then $N_j$ is too high in relation to other ingredients $N_k$ of the system. Then the population $N_j$ is bound to decrease one way or another. As long as there are energy densities differences among the constituents of the system or energy density differences with respect to the surroundings, the system will evolve to decrease free energy, *i.e.*, to increase entropy via diverse processes.

Obviously the mere thermodynamic driving force does not result in evolution but it takes also mechanisms to conduct energy. The equation 2 contains the vital kinetics that is understood by many models of chemical evolution important for life to emerge[39,40,41]. The kinetic rates[9]

$$\frac{dN_j}{dt} = r_j \frac{A_j}{RT} = -\sum_k \frac{dN_k}{dt} \qquad (3)$$

are proportional to the thermodynamic driving forces to satisfy the balance equation. In other words energy and momentum are conserved in the reactions[33]. The coefficient $r_j > 0$ depends on the mechanisms that yields $N_j$. According to the self-similar thermodynamic description each mechanism is a system in itself. For example, an enzyme is a catalytic mechanism that has resulted from a folding process preceded by a chemical synthesis, both evolutionary courses themselves. The coefficient is a constant as long as the mechanism is stationary, *i.e.*, not evolving itself further. When Eq. 3 is inserted to Eq. 2, it is indeed apparent from the quadratic form that $dS/dt \geq 0$. The familiar approximations of the kinetic equation (Eq. 3) are the mass-action law[42] and logistic equations[43] that picture concentrations $N_j$ as motive forces and muddle energetics in variable reaction rates. As a result of using these approximate models that do not spell out free energy as the driving force, kinetics and thermodynamics appear inconsistent with each other. Consequently, thermodynamics seems not sufficient for outlining evolutionary courses and various kinetic scenarios acquire additional emphasis[44,45].

Thermodynamic value of an energy transduction mechanism is only in its ability attain and maintain high-entropy states by energy conduction. The thermodynamic theory is unarmed to say specifically which mechanisms might appear but once some have emerged, their contribution to the reduction of free energy is evaluated according to Eq. 2. Under the energy influx from surroundings the rates of reactions $r_j$ in Eq. 3 are very important because the high-entropy non-equilibrium concentrations compounds and populations of species are constantly replenished by dissipative regeneration. Even a small advantage will accumulate rapidly as an increased flow directs to increase further the population of superior transduction mechanism. This is also known as the constructal law[46].

When some novel compounds happened to appear in the primordial system due to random variation in chemical syntheses, some of them may have possessed some elementary catalytic activity. Even slightly higher rates $r_j$ provided by the emerging catalytic activity were very important to attain more probable non-equilibrium states. They allowed to diminish faster the energy difference between the chemical system and its high-energy surroundings (*e.g.* due to the sunlight). The *dS*/*dt* rate criterion will naturally select faster and faster mechanisms as well as those mechanism that recruit more and more matter and energy from the surroundings to the natural process. Therefore, any primordial energy transduction mechanism that was just slightly faster that its predecessor gained ground. The primitive chemical evolution took the direction of *dS*/*dt* > 0 just as the sophisticated evolution does today. Indeed, contemporary catalyzed reactions contribute to entropy by rapidly producing diverse entities that then interact with each other within their lifetimes, *i.e.*, act as catalysts themselves.

According to the thermodynamics of open systems every entity, simple just as sophisticated, is considered as a catalyst to increase entropy, *i.e.*, to diminish free energy. Catalysis calls for structures. Therefore the spontaneous rise of structural diversity is inevitably biased toward functional complexity to attain and maintain high-entropy states. This quest to level differences in energy by transduction underlies the notion that evolution is progress. Once all differences in energy densities ($\Sigma\mu_k - \Delta Q_{jk} + \mu_j$) have been abolished, the system has reached a stationary state $S_{max}$ and evolution *dS*/*dt* = 0 has come to its end. At this maximum-entropy stationary state, entities keep interacting with each other but there are no net flows of energy among them and no net fluxes from the surroundings to the system or *vice versa*. Frequent mutual interactions maintain the most probable state by quickly abolishing emerging potential differences. The system is stable against internal fluctuations according to the Lyapunov stability criterion[14,47], however when there are changes in surrounding densities-in-energy, the system has no choice but to adapt to them, *i.e.*, to move by abolishing the newly appeared gradients.

## 5. Steps toward life

The primordial pool, the simple chemical system having some abiotic substances in equilibrium numbers $N_1$ began to evolve when a reaction pathway that coupled external energy, opened up and products $N_{j>1}$ began to form. Then the high surrounding potential began to drain into the system as substrates transformed to products. This raised the overall chemical potential of the system toward that of the high-energy radiation. Free energy kept diminishing and entropy continued to increase when reactions yielded more and more products from substrates. During the natural process the initial equilibrium state was lifted up from equilibrium to the non-equilibrium state by the energy influx. Nevertheless, it is important to keep in mind that all flows of energy were downward and still are from high-energy sources to the repositories lower in energy. According to thermodynamics evolution from the equilibrium to the non-equilibrium was a likely sequence of events, not a miraculous singular event. It is the coupling of external energy that made the evolutionary course probable.

The reasoning that the probable course is governed by conditions is in agreement with Le Chatelier's principle, *i.e.*, the conditions determine the stationary state of a reaction. When the external energy coupled to the reactions, the conditions were in the favor of the non-equilibrium stationary state over the equilibrium state. Conversely, when the external energy was reduced (*e.g.* during night or winter), the non-equilibrium state became improbable. Then the system took a course toward the equilibrium, *e.g.*, by consuming established stocks and even disintegrating prior mechanisms of energy transduction during a prolonged starvation.

Remarkably, the equation 1 has not been known explicitly until recently. Importantly it shows that the non-equilibrium state, supported by the external energy, has higher entropy than the equilibrium state. Thus all systems attempt to move toward more probable state by coupling to sources of external energy. The attempt is successful when there are abundant and versatile ingredients to capture the energy influx. To this end carbon chemistry by its impressive number of combinatorial choices was and still is the treasure trove. It allowed numerous mechanisms to emerge, *e.g.*, due to a random variation in the flows, and to increase energy transduction further by channeling more external energy into the system and dispersing it further within the system. Thus the 2$^{nd}$ law of thermodynamics provides the intrinsic bias for emergence of functional structures to conduct energy. The primordial systems, even without genetic material and mechanisms of replication, were subject to evolutionary forces, *i.e.*, directional energy gradients. In the quest to level differences in energy the primordial energy transduction networks expanded and eventually integrated in the global energy transduction system. Thus, it is accurate to say that there is not only life on Earth but the planet is living[48,32].

The thermodynamic formalism is self-similar. It is applicable to diverse levels of hierarchy including complex biological systems that are results of chemical reactions. Thus the thermodynamic description is not only outlining the primordial course of chemical evolution but reveals the characteristics of contemporary processes as well. The question why life emerged and the question what life is are thus tied together. The natural process that accumulated early functional chemical compounds is the one and the same that today involves complex entities, species. The scale is different and mechanisms are versatile and more effective but the principle is the same.

All organisms assemble via numerous chemical reactions. The increase in numbers is, in the case of complex entities, referred to as proliferation (Fig. 1). According to equation 2, entropy is also increasing when different kinds of products appear until the stationary state is attained. In the case of complex entities this process is usually referred to as differentiation that gives rise to biodiversity. In the case of a single organism the process is called developmental differentiation that ends up to the maturity[49], *i.e.*, the stable maximum entropy state. The equation 2 reveals that entropy is increasing further when more external energy couples to the reactions. This process corresponds to an energy intake, *e.g.*, by photo- and chemosynthesis. Entropy will also increase when the system acquires more matter. It is of course known for a long time that entropy of a larger system is higher than a smaller but otherwise a similar system. When the energy intake involves complex entities it is usually referred to as metabolism that powers natural processes such as growth and expansion.

The aforementioned processes from the elementary level of chemical compounds to complex biological entities at higher and higher levels of hierarchical organization are strikingly similar those that we recognize as the basic biological processes. Yet they were exposed simply by considering probabilities of states accessible for an open system undergoing chemical reactions[9]. Thus it is concluded that life is a natural process. It is consequence of increasing entropy, the quest to diminish free energy with no demarcation between inanimate and animate. According to thermodynamics there was no striking moment or no single specific locus for life to originate but the natural process has been advancing by a long sequence of steps via numerous mechanisms reaching so far to acquire a specific meaning – life.

The outlined course of evolution is understood by thermodynamics as a probable scenario. This statement may be interpreted erroneously to imply that life should exist everywhere but apparently does not. Considering the cosmic background spectrum where the appropriate energy range for the processes referred to as biological spans only a minute band, life is undoubtedly rare but not unnatural. The probability is not an abstract concept but inherently associated with energy (also in the form of matter) as is obvious when $S$ in Eq. 1 is multiplied by $T$ to give the overall kinetic energy within the system[33]. Free energy drives evolution so that kinetic energy balances potential energy and energy in radiation. Probabilities are not invariants but keep changing. When there is not much energy or when there are not mechanisms to couple to external energy or not much ingredients to make energy transduction machinery, evolution will not advance very far. The very same laws of thermodynamics that worked in the primordial world are still working today. For example, when a biological system, *e.g.*, an animal is deprived from energy, *i.e.* food, its existence becomes improbable. Thermodynamics is common sense.

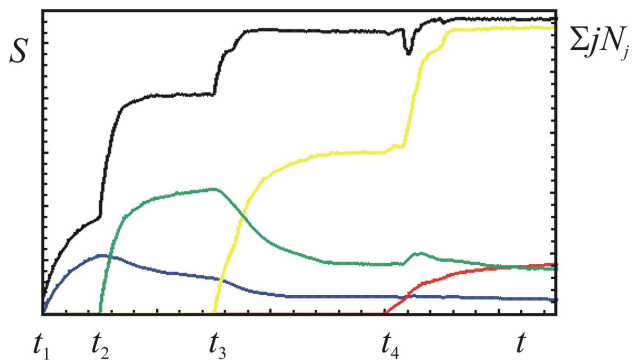

**Fig. 1.** Evolution of a chemical system obtained from a simulation. The simulation was programmed as steps of random syntheses in a for-loop. External energy couples to steps of assembly $N_1 + N_{j-1} \leftrightarrow N_j$ according to Eq. 3 and energy dissipates in dissipative degradations $N_j \leftrightarrow jN_1$. Initially, the system contains only basic constituents in numbers $N_1$. At the time $t = t_1$ a synthesis pathway opens up. Entropy $S$ increases rapidly (black) when matter flows from $N_1$ to new compounds ($j > 1$) in increasing numbers $\Sigma jN_j$ (blue). The growth curve is representative for non-catalyzed reactions. At time $t = t_2$ a second but faster (4x) pathway opens up (green). New kinds of products prompt quickly to the system but soon the system is accumulating them more gradually as energy in the new products becomes comparable to the original but diminishing substrate compounds. The system prompts again when a third pathway punctuates open at time $t = t_3$ yielding catalytic products (yellow) having higher activity with $j$. Later the evolution settles to a new stasis. The form of an autocatalytic growth curve depends on the specific mechanisms. At time $t = t_4$ a fourth pathway opens up (red) yielding products that are capable of slowly recruiting more matter ($N_1$) from outside and maintaining it in the system. As a result the new pathway, even though it is slow, is gaining ground in the overall entropy production. Also with the help of the newest pathway the previously emerged fast catalytic pathway will have more matter to yield even better catalysts to attain higher states of entropy whereas the relative contribution of older slower pathways continues to diminish, eventually facing extinction.

## 6. The equation of evolution

Considering the explanatory power of thermodynamics, it is perhaps surprising that the probable course of evolution cannot be solved and predicted in detail. The fundamental reason is exposed by rewriting the equation 2 for the probability using the definition $S = R\ln P$

$$\frac{dP}{dt} = LP \geq 0 \, ; \quad L = \sum_{j=1} \frac{dN_j}{dt} \frac{A_j}{RT} \, . \tag{4}$$

The equation of motion cannot be solved analytically[9] because the driving forces $L$ keep changing with changing flows. The non-conserved system, summarized by the probablity $P$, is changing because its energy content is either increasing or decreasing. Chemical reactions are endo- or exoergic, *i.e.*, it is imposible for the system to change its state without acquiring or loosing a quantum. In other words, there are no invariants of motion which is the fundamental reason for the unpredictable courses of evolution. New mechanisms accessing new potentials are in turn transformed into new mechanisms that redirect the flows of energy and so on. Even small perturbations in the initial conditions affect the overall course and evolution is per definition chaotic[47].

Despite evolution being non-deterministic its main charateristics are revealed by the equation of motion. Notably when new means appear to conduct energy from plentiful potentials, the probablity will increase rapidly. Then evolution punctuates because suddenly there is much to draw from and thus according to Eq. 3 the rate $dN_j/dt$ is fast. When the supplies narrow, the process slows down. Finally when the net resources have become exhausted, the system settles to a stasis. This characteristic course of punctuations and stases[50] covers both complex animate and simple inanimate systems[51] (Fig. 1). For the large global ecosystem the evolutionary course has taken eons whereas a simple and small system will settle fast to a stasis.

The maximum-entropy steady-state distributions of energy transduction mechanisms, *e.g.*, populations $N_j$ of species that result from natural processes, are characteristically skewed[28,29,30]. The distribution contains relatively few most expensive mechanisms at the top of the energy transduction chain, *i.e.*, food chain. They are thermodynamically expensive hence rare but highly effective in energy transduction. The numerous mechanisms at the intermediate levels are not particularly expensive but altogether conduct most of energy. The most inexpensive entities do not have much of mechanisms and thus they will not contribute much to the overall energy transduction either.

The propagator $L$ in Eq. 4 denotes the energy landscape by tangential vectors that keep changing as energy flows[33]. A coordinate on the manifold of energy densities distinguishes from another coordinate by energy thereby expressing the concept of identity in terms of energy. Therefore evolution as an energy transduction process can be viewed as an energy landscape in a flatting motion. The thermodynamic analysis reveals that the manifold is not preset, *i.e.* deterministic. It is non-Euclidian because the "distances" in free energy are directional (thus not proper distances) and because the "distance" between two energy densities will change when a third density of energy comes within interaction range (thus the triangle inequality need not be satisfied).

## 7. Discussion

To understand origins and evolutions of complex systems, thermodynamics calls our attention not to discard the principle of decreasing free energy equivalent to the principle of increasing entropy. Often the universal thermodynamic principle and the natural selection in the theory of evolution are viewed as opposing forces. This is a misconception. The driving force due to external energy has remained obscure because the equation for the rate of entropy increase (Eq. 2) has been deduced but not derived from the first principle probability calculation[9]. Furthermore, when the entropy concept was formulated by statistical physics, free energy was not recognized as the evolutionary force because it is absent at the equilibrium that was determined mathematically using Lagrange multipliers rather than following the course directed by fading forces. Consequently, the concepts of entropy and order have become mixed with each other. Owing to the confusion it has become accustomed to say that living systems would export entropy to maintain their internal high degree of order. The objective is not to maintain order but to employ orderly energy transduction machinery to diminish energy gradients. The vital orderly mechanisms of energy transduction are not low in entropy, *i.e.* improbable, when being parts of an external energy-powered system. It is emphasized that entropy increases when differences in energy diminish whereas disorder, more precisely decoherence, increases during isergonic processes due to stochastic exchange of quanta. Indeed the pedagogical cliché of equating entropy with disorder is unnecessarily confusing and ultimately wrong[52]. The common misconception that entropy of a living system could possibly decrease at the expense of entropy increase in its surroundings does in fact violate conservation of energy. It is possible, although statistically unlikely, that entropy of a system and its surroundings both would decrease. This means that energy would transiently flow upwards from a low to a high-density. Thus the $2^{nd}$ law of thermodynamics and the theory of evolution by natural selection are not opposing but one and the same imperative. There is no demarcation line between animate and inanimate.

The natural selection by the rate of entropy increase among alternative ways, *i.e.*, mechanisms to conduct energy is the self-consistent and universal criterion of fitness. In the primordial world any mechanism, irrespective how simple or elementary, did do to move toward more probable states. Primordial catalysts, perhaps yielding only minute rate enhancements, could just have been the compounds themselves. Later, when other ways but faster opened up they were employed to reach states that were even higher in entropy. Thus evolution is tinkering[53], and there might be only very little clues left to track down specific chemical reactions that began to increase the energy content of matter on Earth by coupling to high-energy flux from Sun. Nevertheless, the emergence of systems with increasingly higher degrees of standards such as chirality in biological macromolecules and common genetic code can be recognized as sign posts of evolution. We see nothing of these slow changes in progress, until the hand of time has marked the long lapses of ages[1].

When a system cannot access more matter or energy, the rates of energy transduction may still continue to improve to reach higher states of entropy. The rates of entropy increase are relative to one another. When ingredients are intrinsically difficult to recruit to the natural process, even a slow process is better than nothing. The $dS/dt$ rate is a blind but highly functional criterion. Over the eons rates have improved over and over again to result in, *e.g.*, efficient cellular metabolism and ecosystem food web. Today catalyzed kinetics is so ubiquitous characteristic of life that it is easily regarded as a profound cause rather than being a consequence of the principle of increasing entropy by decreasing gradients in energy. The $dS/dt$ rate criterion guarantees that only those among diverse entities that are capable of contributing to entropy are maintained in the system, *i.e.*, will survive. The rate of entropy increase as the selection criterion resolves the circular argument: fitness marks survival – survival means fitness. Natural selection by the entropy increase rate may at first appear merely as a conceptual abstraction or an oversimplification of reality. Indeed it may be difficult to recognize the increase of entropy, equivalent to the decrease of free energy, as the common motive among many and intricate contemporary mechanisms of life. However intricacies and complexities are in the machinery, not to be confused with the universal objective.

The principle of increasing entropy explains why matter organizes in functional structures and hierarchies. The order and complexity in biological systems has no value as such. Mechanisms and structures are warranted only by their energy transduction, *i.e.*, ability to attain and maintain high-entropy states. A system cannot become larger than the one where its entities still reach to interact with each other. For example, molecules that are results of endoergic external energy powered reactions, are bound to break down and thus they may take part only in the reactions that they will reach within their lifetimes. Further entropy increase may take place when systems themselves become entities of a large system at a higher hierarchical level with a larger range of interactions. For example, molecules are entities of systems known as cells that are entities of organisms and so on. The principle $dS > 0$ is also the universal condition of integration. An organization will form when entropy increases more than can be achieved by entities as systems interacting with their surroundings independently. Some organisms, *e.g.*, yeast exemplify the thermodynamic principle by switching between uni- and multicellular modes of organization depending on surrounding supplies, the potential energy gradients. Thus a hierarchical organization is just a mechanism among many others to conduct energy.

According to thermodynamics mechanisms are consequences of the natural process, not conditions for life to emerge. There is no requirement for an autocatalytic self-replicating molecule being assembled by a fortuitous event and susceptible to mutations for natural selection to operate on it. This is in agreement with the notion "metabolism first" however without incentive to discover a specific, vital mechanism. There is no problem for evolution to take its direction. It is always down along the energy gradients. The role of heredity and information is not overlooked by the thermodynamic formalism either. It is incorporated in the evolutionary processes as mechanisms. The physical view of information gives understanding, *e.g.*, to its dispersal in genomes.

The unifying view of thermodynamics captures courses and distributions of matter with no demarcation line between living beings and inanimate. Stochastic processes act on *all* matter and to put it in motion toward increasing entropy. The result is evolution, *i.e.*, a series of steps from one state to another to lower potential energy differences. Earth, our home, is in between the huge potential energy difference due to the hot Sun and the cold space. Biota has emerged, as integrated in atmospheric and processes of geosphere, to diminish the difference by energy transduction. The theory of evolution by natural selection formulated in thermodynamics roots biology via chemistry to physics to widen contemporary discourse on fundamentals of evolution and emergence of life.

**Acknowledgments**

We thank Christian Donner, Sedeer El-Showk, Mikael Fortelius, Carl Gahmberg, Salla Jaakkola, Kari Keinänen, Liisa Laakkonen, Martti Louhivuori, Kaj Stenberg, Mårten Wikström and Peter Würtz for enlightening discussions and valuable comments.